\documentclass[a4paper,11pt]{article}
\usepackage{jcappub}
\usepackage{graphicx}
\usepackage{caption}
\usepackage{subcaption}
\usepackage{xcolor}

\newcommand{\gd}{$g_{10}$}

\definecolor{darkblue}{rgb}{0,0,0.5}
\definecolor{darkgreen}{rgb}{0.1,0,0.3}
\definecolor{darkred}{rgb}{0.6,0,0}

\title{New axion and hidden photon constraints from a solar data global fit}

\author[a,1]{N. Vinyoles,\note{Corresponding author.}}
\author[a]{A. Serenelli,}
\author[b,c]{F.L. Villante,}
\author[d]{S. Basu,}
\author[e,f]{J. Redondo}
\author[a]{and J. Isern}

\affiliation[a]{Institute of Space Sciences (CSIC-IEEC),\\Facultad de Ciencies, Campus UAB, Torre C5 parell 2, 08193 Bellaterra, Spain}
\affiliation[b]{Dipartimento di Scienze Fisiche e Chimiche, Universit\`a dell'Aquila, \\ 
I-67100 L'Aquila, Italy}
\affiliation[c]{Istituto Nazionale di Fisica Nucleare (INFN), Laboratori Nazionali del Gran Sasso (LNGS), \\ I-67100 Assergi (AQ), Italy}
\affiliation[d]{Department of Astronomy, Yale University, \\ PO Box 208101, New Haven, CT 06520, USA}
\affiliation[e]{Departamento de F\'isica Te\'orica, Facultad de Ciencias, Universidad de Zaragoza,  \\
50009 Zaragoza, Spain}
\affiliation[f]{Werner-Heisenberg-Institut, Max-Planck-Institut fur Physik, \\ F\"ohringer Ring 6
80805 M\"unchen, Germany}

\emailAdd{vinyoles@ice.csic.es}
\emailAdd{aldos@ice.csic.es}
\emailAdd{francesco.villante@lngs.infn.it}
\emailAdd{sarbani.basu@yale.edu}
\emailAdd{jredondo@unizar.es}
\emailAdd{isern@ice.csic.es}

\abstract{We present a new statistical analysis that combines helioseismology (sound speed, surface helium and convective radius) and solar neutrino observations (the $^8$B and $^7$Be fluxes) to place upper limits to the properties of non standard weakly interacting particles. Our analysis includes theoretical and observational errors, accounts for tensions between input parameters of solar models and can be easily extended to include other observational constraints. We present two applications to test the method: the well studied case of axions and axion-like particles and the more novel case of low mass hidden photons. For axions we obtain an upper limit at $3\sigma$ for the axion-photon coupling constant of $g_{a\gamma}\,<\,4.1 \cdot 10^{-10} \rm{GeV^{-1}}$. For hidden photons we obtain the most restrictive upper limit available accross a wide range of masses for the product of the kinetic mixing and mass of $\chi m < 1.8 \cdot 10^{-12} \rm{eV}$ at $3\sigma$. Both cases improve the previous solar constraints based on the Standard Solar Models showing the power of using a global statistical approach.}


\arxivnumber{1501.01639}

\begin{document}
\maketitle
\flushbottom

\section{Introduction and motivation}
\label{sec:intro}

The Standard Model of Particles (SM) is not a complete theory, and extensions are necessary to address some of the most pressing open questions in fundamental physics. Among others, some of these are: nature of dark matter, matter-antimatter asymmetry in the Universe, origin of neutrino masses, strong CP-violation problem. In order to solve these problems, physics beyond the SM is needed and the existence of new particles and/or non-standard properties of known particles are generally invoked. 

Stellar physics, while far from being a closed subject, is a mature discipline and it provides an accurate understanding of the internal structure of stars and their evolution. The sheer mass and size, the extreme conditions reigning in stellar interiors and, in many cases, the extremely long lifetimes that allow to \emph{integrate} small effects over a very long time make stars appealing as laboratories for particle physics under conditions not reproducible anywhere else.

In particular, the existence of weakly interacting particles beyond the standard model  or of standard particles with non-standard properties can modify the internal structure and its evolution in different ways. For example, weakly interacting massive particles (WIMPs), too massive to be created inside stars, can be accreted from the dark matter halo of the Milky Way and contribute to the transport of energy inside stars \cite{spergel1985,taoso2010,vincent2014} and, in the case of self-annihilating particles act as a localized energy source. Weakly interacting light particles, on the other hand, can be thermally produced in the stellar interiors and can easily escape due to their large mean free path and act as energy sinks. The increased rate of energy-loss then changes the internal structure of stars and also modifies evolutionary timescales. The production rate of each particle depends differently on stellar conditions. Consequently, the impact of different particles on stellar structure will depend on the class of stars and on the evolutionary state. Several works have given constraints on the properties of the weakly interacting light particles using different stellar objects. Some examples are: globular clusters, where the luminosity of stars at the tip of the red giant branch \cite{raffelt90,raffelt95,viaux2014} and the lifetime of horizontal branch stars \cite{part-raffelt88, an2013, solaredondo2013, ayala2014} can be used to give constraints on axion (e.g. strength of coupling to photons and electrons) and neutrino (magnetic dipole moment) properties; intermediate mass stars for which sufficient cooling during the helium-core burning phase can make the so-called blue loop disappear, thus contradicting the observation of Cepheids~\cite{Friedland:2012hj}; white dwarf stars, where the anomalous cooling induced by axions and axion-like particles shortens their cooling timescale affecting both their luminosity function (LF)~\cite{Raffelt:1985nj,isern2008,miller2014,Dreiner:2013tja} and pulsation properties \cite{Isern:1992gia,Isern:2010wz,2012ASPC..462..533C}; neutron stars, where additional axion energy losses~\cite{Keller:2012yr,Umeda:1997da} can even lead to observable deviations of surface cooling in real time~\cite{Leinson:2014ioa}; SN1987A, where the extra cooling of the proto-neutron core would shorten the neutrino pulse~\cite{Raffelt:1987yt,Turner:1987by,Mayle:1989yx,Mayle:1987as} and the axions produced could produce a prompt gamma ray signal \cite{Turner:1987by,Payez:2014xsa}. These ideas have been widely discussed in the literature, the most comprehensive reference being the seminal book by G. Raffelt \cite{book-raffelt96}, which can be complemented with recent topical reviews on QCD axions~\cite{Raffelt:2006cw,Agashe:2014kda}, axion-like particles~\cite{Carosi:2013rla} and hidden photons~\cite{solaredondo2013}.

Many studies have focused on using the Sun for setting limits on the properties of different types of particles; we review some of them below. The Sun is by far the best-known star. The solar structure, revealed by helioseismology and solar neutrinos, is well determined, and accurate solar models give us information about the past, present and the future of the Sun \cite{serenelli11}. While in some cases (e.g. axions) the most restrictive limits are not inferred from solar studies, the Sun remains the most useful benchmark for testing and validating both stellar models and, as it is partly the case of the present paper, different statistical approaches to constrain particle properties. Also, it is important to keep in mind that CAST \cite{CAST} and the forthcoming IAXO \cite{IAXOweb,IAXO} are experiments specifically designed to detect exotic particles directly from the Sun, so having predictions of expected solar fluxes for exotic particles remains an important aspect to consider. 

Solar constraints on particle properties have been generally derived from applying limits to variations of either neutrino fluxes \cite{part-schlattl99,part-gondolo09} or the sound speed profile derived from helioseismology~\cite{part-schlattl99}. However, a systematic approach aimed at combining different sources of data accounting in detail for observation and theoretical errors is badly missing in the literature. Here, we try to supply such a tool.

The initial goal of this work is to extend the general statistical approach presented in \cite{villante2014} to  constrain properties of particles  (e.g. mass, coupling constant) making the best possible use of the available information on the Sun, both observational and theoretical. To do so we use the helioseismic data combined with the neutrino fluxes in a statistical approach that includes the theoretical and observational uncertainties and takes into account possible tensions among data and solar model input parameters. We then derive solar limits for the well-studied hadronic axions --to gauge the performance of our statistical approach-- and for the more novel case of hidden photons for which the Sun sets the most restrictive limits on the kinetic mixing parameter for small hidden photon masses, $m\lesssim $eV.

Axions are light pseudoscalar particles that were introduced by the Peccei-Quinn \cite{pecceiquinn77} solution to the strong CP problem. They are very light particles and interact with ordinary particles much like neutral mesons ($\pi^0,\eta$,...) but with coupling strengths vastly weaker, and some model dependencies. The most relevant for astrophysics is whether axions couple with electrons with similar strength than to nucleons, because this coupling tends to be very efficient to produce axions in stellar environments.  The DFSZ model~\cite{dine1981} as any axion model embedded in a GUT are examples of axion models with tree-level axion coupling. Here, the main axion production mechanisms are the ABC proceses: axio-recombination, bremsstrahlung and Compton~\cite{redondo13}.   
In the KSVZ model (hadronic axions)~\cite{kim1979,shifman1980} this coupling is absent at tree level and the relevant axion coupling is to two photons. Here, the Primakoff effect is the mechanism that converts photons into axions in the  presence of electric or magnetic fields. The inverse Primakoff effect is used in helioscopes like CAST and SUMICO (and the proposed IAXO) to convert solar axions into detectable X-rays and constrain the solar axion flux.  

The phenomenology of axions is extended to axion-like particles (ALPs) in a completely straight-forward way. ALPs are also 
bosons with a two-photon coupling, but their mass and interaction strength are in principle unrelated, unlike the axion case. 
These particles arise copiously in string compactifications (a string axiverse~\cite{Dias:2014osa,Cicoli:2012sz,Arvanitaki:2009fg}) , where O(100) ALPs can be generically invoked, 
although the number of light enough particles to be relevant for stellar physics is completely model-dependent. 
ALPs can be behind some very interesting hints recently pointed out in astrophysics like the transparency of the universe to gamma-rays~\cite{Meyer:2013pny}, the 3.55 keV line~\cite{Higaki:2014zua,Jaeckel:2014qea,Alvarez:2014gua}and the soft excess of the coma cluster ~\cite{Angus:2013sua}.

In this paper, we are primarily interested in the case of hadronic axions because the Sun is one of the most sensitive environments to look for their effects.  Solar constraints are even more important for the case of ALPs, where the SN1987A constraint is absent in generic models (ALPs do not generically have large couplings to nucleons but axions do). We shall not endeavour constraining the axion-electron coupling because limits from white dwarves and red-giant stars are comparatively much stronger. 

Until now, several works have provided constraints to the axion-photon coupling constant $g_{a\gamma}$ using the variations that axions produce on  helioseismologic quantities or neutrino fluxes. An upper limit for $g_{a\gamma} = 10 \cdot 10^{-10} {\rm GeV^{-1}}$ is found by setting a limit to the deviation that axions can impart to the solar model sound speed at a given depth in the Sun \cite{part-schlattl99}. This work  also gives values for the solar neutrino fluxes depending on axion emission and \cite{part-gondolo09} uses this relation to give a more restrictive constraint of $g_{a\gamma} = 7 \cdot 10^{-10} {\rm GeV^{-1}}$ at a $3\sigma$ confidence level using the observed $\Phi(^8{\rm B})$ solar neutrino flux measured by the SNO experiment (see \cite{sno2013} for the global analysis of the three SNO phases). In \cite{maeda13} they  construct the so-called seismic models (non-evolutionary solar models constructed in ad-hoc manner to reproduce the sound speed derived from helioseismology), with different values for the axion-coupling constant and obtain an upper limit of $g_{a\gamma}= 2.5 \cdot 10^{-10} {\rm GeV^{-1}}$ by comparing the predicted $\Phi(^8{\rm B})$ with the experimental result and using  $1\sigma$ uncertainties. For the mass range $m_a \leqslant 0.02\, \rm{eV}$, the most restrictive limit comes from the helioscope CAST with $g_{a\gamma} < 0.88 \cdot 10^{-10} \rm{GeV^{-1}}$ \cite{andriamonje2007}. The future helioscope IAXO should improve these results, as it is expected to reach sensitivities to the axion-coupling constant 1 to 1.5 orders of magnitude better than CAST \cite{IAXOweb}. 

Hidden photons, our second case of study, are vector bosons that couple weakly via kinetic mixing with standard photons. The kinetic mixing is represented by the parameter $\chi$ and  together with the hidden photon mass, $m$, are the parameters that must be constrained (see  \cite{solaredondo2008,jaeckel2010,jaeckel2013,solaredondo2013} and references therein). 
In fact, as described in Section~\ref{sec:hiddenmodels}, solar constraints are mostly sensitive to the product $\chi m$ if $m\,\lesssim\,{\rm eV}$.

Hidden photons can only be produced from photon $\leftrightarrow$ HP oscillations, which are affected by the photon refraction in the solar plasma. The oscillations are resonant when the hidden photon and photon dispersion relation match and this happens differently for transversely polarized photons and longitudinal excitations (L-plasmons). Resonant emission in the Sun is possible for hidden photon masses below $\sim$0.3 keV, the highest plasma frequency $\omega_{\rm p}$, which is then also the highest produced photon effective mass.
The emission of L-modes is more important for low hidden photon masses (below $m\sim$eV)~\cite{an2013,solaredondo2013}, for which  resonant conditions happen all through the solar interior (each region of the Sun emits L-HPs with energy equal to the local plasma frequency). This is the case of interest in this paper. 
Resonant emission of T-modes dominates the energy loss in the hidden photon mass range $\sim$ eV-0.3 keV and it is localized in a narrow spherical shell of the solar interior for which the hidden photon mass matches the plasma frequency $m\simeq \omega_{\rm p}$. This case is very interesting too, but our solar model codes need to be tuned for such highly localized sources, a task that we plan to endeavor in a forthcoming publication. For even higher masses, arguments from horizontal branch stars can give better constraints because, owing to the higher plasma frequencies, resonant production extends to higher masses, and when it is not possible the higher temperatures ensure the existence of photons of sufficient energy to account at least for the hidden photon mass. Therefore, it is not crucial to reexamine this regime. 

This paper is structured in the following way. In Section~\ref{sec:models} we present the  theoretical and experimental aspects of this work: solar models and data used and their respective treatment of uncertainties. In Section~\ref{sec:methods} we introduce the statistical method, in Section~\ref{sec:results} we describe our main results and present new upper bounds to the axion mass and kinetic coupling of hidden photons. In Section~\ref{sec:discussion} we compare our work to previous results and draw some conclusions and a short summary in Section~\ref{sec:summary}.

\section{Models and Data}
\label{sec:models}

\subsection{Standard Solar Models}

In this work  we use standard solar models (SSMs) as reference models. SSMs are calibrated to match the present-day solar radius ${\rm R_\odot}=6.9598\cdot10^{10}$~cm, luminosity ${\rm L_\odot}=3.8418\cdot10^{33}$~erg~s$^{-1}$ and surface metal-to-hydrogen ratio ${\rm (Z/X)_\odot}$. The choice of this last constraint is critical because it essentially determines the distribution of metals in the entire solar structure and it has been the subject of much discussion over recent years in the context of the \textit{solar abundance problem} \cite{basu2004b,bahcall2005a,serenelli09,guzik2010}.

 One frequently adopted solar photospheric composition is that provided by Grevesse and collaborators (hereafter GS98) \cite{gs98ab}, based on 1-D solar atmosphere models, from which ${\rm (Z/X)}_\odot = 0.0229$. SSMs based on GS98 composition are in good agreement with helioseismic observables and solar neutrino fluxes determinations. A more recent compilation of the solar photospheric abundances has been provided by  Asplund and collaborators \cite{agss09ab} (hereafter AGSS09). It is based on a more accurate analysis with new 3-D hydrodynamical  models of the solar atmosphere that reproduces very well the observed solar atmosphere properties (granulation, convective velocities and limb darkening among others). As a result of these and other improvements in solar spectroscopic analysis, they obtain a much lower surface metallicity $(Z/X)_\odot=0.0178$ that leads 
to SSMs in strong disagreement with the solar structure inferred from helioseismology. More quantitatively, SSMs computed with AGSS09 surface composition disagree with the observed value of the convective radius ${\rm R_{CZ}}$ and surface helium abundance ${\rm Y_{S}}$ at $2.5 \sigma$ and $3.4 \sigma$ while when using GS98 composition one obtains agreement within  $0.2 \sigma$ and $1.1 \sigma$, respectively.  For the sound speed profile the situation is similar with average rms deviations from helioseismic determinations a factor 4 worse for the AGSS09 models compared with the GS98 ones \cite{serenelli11,villante2014}.

In \cite{cd2009}, and more recently in \cite{villante2014}, it was pointed out that current seismic and neutrino observables are only sensitive to the actual opacity profile\footnote{The two exceptions are the adiabatic index $\Gamma_1$, see \cite{lin2007}, that can be used together with the equation of state, to determine the metallicity of the solar envelope and the CNO neutrino fluxes that can be used to determine the sum of the carbon and nitrogen content of the solar core.} that defines the temperature stratification of the Sun. The opacity profile is determined by a set of  radiative opacity calculations (eg. OP \cite{op2005},OPAL \cite{opal1996}) and the solar composition (eg. GS98 or AGSS09). While it is not possible to separate these two contributions, the solar opacity profile is well constrained by helioseismic and solar neutrino experiments \cite{villante2014}. 
%Then, for our purpose we look for a model with an opacity profile close to the observaitions. 
The production of the exotic particles considered in this paper and their effects on solar structure depend on the thermal structure of the Sun, not on its detailed composition. This is evident by considering that energy-loss rates depend mainly on temperature and density. Therefore, the fact that the present observational constraints determine very well the thermal stratification of the Sun allow us to use these models to test the impact of exotic particles regardless of the detailed knowledge of the composition of the Sun.

We have computed all our models by using GARSTEC (GARching STEllar Code) \cite{GARSTEC}. The physics used in the calculation of SSMs is as follows: the 2005 update to the OPAL equation of state \cite{opal2002}, OP opacities \cite{op2005} complemented at low temperatures with molecular opacities \cite{ferguson2005} and nuclear rates from Solar Fusion II \cite{adelberger2011}. All models include microscopic diffusion as described in \cite{thoul1994}. For more details about the SSMs used on this work see \cite{serenelli11}.

\subsection{Models with axions}
\label{sec:axionmodels}

The production of axions in the Sun occurs via the Primakoff effect, i.e. axions are produced by the conversion of photons in the electric field of nuclei and electrons with the interaction Lagrangian $\mathcal{L}_{a\gamma} = g_{a\gamma} \textbf{B} \cdot \textbf{E} \,a$, where $a$ is the axion field. Thus, constraints can be placed on axion-photon coupling constant $g_{a\gamma}$. The energy-loss rate per unit mass $\epsilon_{a\gamma}$ is given by \cite{part-schlattl99}: 
%equations \ref{eq:primakoff}-\ref{eq:kappa}.
\begin{equation}
\epsilon_{a\gamma}=\frac{g_{a\gamma}^{2}}{4\pi}\frac{T^7}{\rho} F(\kappa^2),
\label{eq:primakoff}
\end{equation}
where T is the temperature, $\rho$ the density and $F(\kappa^2)$ is a dimensionless function describing screening effects given by:
\begin{equation}
F(\kappa^2)=\frac{\kappa^2}{2 \pi^2} \int_0^\infty dx \frac{x}{e^x-1} \left[ (x^2 + \kappa^2) \rm{ln} \left( 1 + \frac{x^2}{\kappa^2}\right) - x^2 \right].
\end{equation}
Here, the parameter $\kappa$ is defined as: 
\begin{equation}
\kappa^2=\pi\alpha\frac{n_B}{T^3} \left(  {\rm Y_e} + \sum_j {\mathcal Z}^2_j \, {\rm Y}_j \right).
\label{eq:kappa}
\end{equation}
where $\alpha$ is the fine structure constant, $n_B$ the baryon density, ${\rm Y_e}$ the electrons per baryon, 
${\rm Y}_j={\rm X}_j/{\mathcal A}_j$ and ${\rm X}_j$, ${\mathcal A}_j$ and ${\mathcal Z}_j$ represent, respectively, the mass fraction, atomic weight and atomic number of the nuclear species $j$. 
For solar conditions the function $F(\kappa^2)$ can be approximated by \cite{part-schlattl99}, 
\begin{equation}
F(\kappa^2)=1.842(\kappa^2/12)^{0.31}.
\end{equation}

Solar models have been computed for different $g_{a\gamma}=~g_{10}~\cdot~10^{-10}$~${\rm GeV^{-1}}$ values, where  $g_{10}$ spans the range from 0 to 20 with an interval of $\Delta {g_{10}}=1$. Two sets of models have been computed for the GS98 and AGSS09 solar composition as reference compositions. However, following \cite{villante2014}, we treat the solar composition as free parameters in our analysis, as discussed later in Sect.~\ref{sec:methods} in more detail.

%We have computed two different sets of solar models varying $g_{a\gamma}$ and using different surfaces composition (GS98 or AGSS09). Each set is composed of 21 models with $g_{a\gamma}=~g_{10}~\cdot~10^{-10}$~${\rm GeV^{-1}}$, where $g_{10}$ ranges from 0 to 20 with an interval of $\Delta {g_{10}}=1$.

\subsection{Models with hidden photons}
\label{sec:hiddenmodels}

The production of hidden photons ($\gamma'$) in the interior of the Sun can be seen as $\gamma - \gamma'$ oscillations, described by the Lagrangian 
\[ 
\mathcal{L}=-\frac{1}{4}A_{\mu\nu}A^{\mu\nu}-\frac{1}{4}B_{\mu\nu}B^{\mu\nu}+\frac{m^2}{2}B_{\mu}B^{\mu}-\frac{\chi}{2} A_{\mu\nu}B^{\mu\nu}, 
\]
where $A_{\mu\nu}$ and $B_{\mu\nu}$ are the field strengths of the photon and hidden photon field, $A_\mu$ and $B_{\mu}$ respectively.  The energy loss rate per unit mass $\epsilon_{hp}$ is calculated by using the approximation presented in \cite{solaredondo2013}. We only take into account the dominant process of resonant emission of longitudinal hidden photons (details in \cite{solaredondo2013,solaredondo2008}) for which the energy loss rate is given by
\begin{equation}
\epsilon_{hp}=\frac{\chi^2m^2}{e^{\omega_P/T}-1}\frac{\omega_P^3}{4\pi}\frac{1}{\rho}
\label{eq:HP}
\end{equation}
where $\chi$ is the kinetic mixing parameter, $m$ the mass of the hidden photon and $\omega_P$ is the characteristic plasma frequency.  Typical values in the solar center are $\omega_P \sim 0.3~\rm{keV}$ and  $T \sim 1~\rm{keV}$. By expanding the exponential, it can be seen that the temperature dependence of the energy-loss rate is linear to first order in $\omega_P/T\sim 0.3$.  In principle, a threshold factor $\sqrt{1-(m/\omega_P)^2}$ should be included in Eq.~(\ref{eq:HP}), but this term is completely irrelevant for the range of masses considered in this paper, i.e. $m\lesssim$ eV. As a consequence, the energy loss rate $\epsilon_{hp}$  depends only on the product $\chi m$ which is the quantity that can be constrained by solar data.   For hidden photons we consider $\chi m$ ranging from 0  to $8~\cdot~10^{-12}~\rm{eV}$  with an interval $\Delta\chi m =1~\cdot~10^{-12}~\rm{eV}$. As for axions, models with both GS98 and AGSS09 composition as reference have been computed, but the complete treatment of composition is given in Sect.~\ref{sec:methods}.

\subsection{Observables}\label{sec:observables}

The observable quantities used in this analysis are the boron and beryllium neutrino fluxes $\rm{\Phi(^8B)}$ and $\rm{\Phi(^7Be)}$, the convective envelope properties inferred for helioseismology, i.e. the surface helium abundance $\rm{Y_S}$ and the convective radius $\rm{R_{CZ}}$, and the solar sound speed profile $\rm{c_s}(r)$.

The  solar neutrino fluxes determined from experimental data are taken from \cite{serenelli11}, and have been derived using all available experimental neutrino data using the method described in  \cite{bahcall2003}.

The observational values and errors together with the corresponding SSM predictions of the solar neutrino fluxes and convective envelope properties are specified in Table~\ref{tab:obs_value}.
%We adopt the same model errors as in \cite{villante2014}.
\begin{table}[h]
\begin{center}
\begin{tabular}{ c c c c c}
Q & AGSS09 & GS98 & Observables & Ref. \\
\hline
$\rm{Y_S}$ & $0.232 (1 \pm 0.013)$ & $0.243 (1 \pm 0.013)$ &$0.2485 \pm 0.0035$& \cite{basu2004b}\\
$\rm{R_{CZ}/R_{\odot}}$ & $0.7238 (1 \pm 0.0033) $&$0.7127 (1 \pm 0.0033)$& $0.713 \pm 0.001$& \cite{basu1997}\\
$\rm{\Phi(^7Be)}$ & $4.56 (1  \pm 0.06) $ & $5.00 (1 \pm 0.06)$& $4.82(1^{+0.05}_{-0.04})$ & \cite{serenelli11,bahcall2003}\\
$\rm{\Phi(^8B})$ & $4.60 (1 \pm 0.11)$ & $5.58 (1 \pm 0.11)$ &$5.00(1 \pm 0.03)$& \cite{serenelli11,bahcall2003}\\
\hline
\end{tabular}
\caption{\small The first two columns show the SSM predictions for GS98 and AGSS09 composition and the corresponding theoretical uncertainties \cite{villante2014}. Note that theoretical errors do not include uncertainties due to the solar composition. The third column summarizes the observational values and errors. Neutrino fluxes are in $10^9 \rm{cm^{-2} s^{-1}}$ for the $\rm{\Phi(^7Be)}$ and $10^6 \rm{cm^{-2} s^{-1}}$ for $\rm{\Phi(^8B)}$.
\label{tab:obs_value}}
\end{center}
\end{table}

For the sound speed profile, we take the 30 radial points $r_i$ in the range $0.05 < r_i/{\rm R_\odot} < 0.80$ given by \cite{basu09} and use the helioseismic determination of the sound speed ${\rm c}_i$ in those points as our observable quantities. The inversion of the sound speed profile has a dependence on the reference model \cite{basu2000} that, however mild, is better to take into consideration. For this reason, the solar sound speed profile has been derived using helioseismic data independently for each model considered in this work, i.e. using consistently each model for the helioseismic inversion. Inversion has been done using the SOLA inversion technique. The adopted frequencies are the BISON-13 dataset complemented with data of MDI, GOLF and IRIS. More details on both the frequency dataset and inversion technique are given in \cite{basu09}.

\subsection{Uncertainties}
\label{Uncert}

There are two different error sources in this work: the fully correlated theoretical errors, induced by uncertainties in the input parameters for solar model construction, and the observational errors associated to the helioseismic data and the experimental solar neutrino fluxes. 

The errors on theoretical predictions are calculated by propagating the uncertainties of SSM input parameters $I$. 
As in \cite{villante2014}, the input parameters are: the age of the Sun, diffusion coefficients, luminosity, opacity and astrophysical factors of relevant nuclear reactions ($\rm{S_{11}}$, $\rm{S_{33}}$, $\rm{S_{34}}$, $\rm{S_{17}}$, $\rm{S_{e7}}$, $\rm{S_{1,14}}$). We do not include the chemical composition uncertainty because we consider the abundances of volatile and refractory elements as free parameters in the fit, as it is explained in Sect.~\ref{sec:methods}. For each observable quantity $Q$, we calculate the fractional change $C_{Q,I}$ that is obtained when the fractional variation $\delta I$ is applied to the input $I$. 
This coefficient is obtained as
\begin{equation}
C_{Q,I}=B_{Q,I} \, \delta I
\end{equation} 
where
\begin{equation}
B_{Q,I}=\frac{\partial \rm{ln} {Q}}{\partial \rm{ln} {I}}. \label{eq:bqi}
\end{equation}
The values of the power-law exponents $B_{Q,I}$ and of the input parameters uncertainties $\delta I$ are given in \cite{serenelli13} and \cite{villante2014}.
The total theoretical error for $Q$ is found by combining in quadrature all the input error contributions (see \cite{villante2014} and \cite{serenelli09} for details):
\begin{equation}
\sigma_{Q,\rm th}^2=\sum_{I}C^2_{Q,I}.
\label{eq:errcontr}
\end{equation}
Errors for the theoretical neutrino fluxes and convective envelope properties
are given in the first two columns of Table~\ref{tab:obs_value}.

The observational uncertainties $U_Q$ are treated as uncorrelated. For neutrino fluxes and convective envelope properties, they correspond to the observational errors summarized in 
the third column of Table~\ref{tab:obs_value}.
For the sound speed, a relevant contribution is provided by errors in the inversion of helioseismic data. Therefore, at each radial point $r_i$ we calculate the total uncertainty as:
\begin{equation}
U_{{\rm c}}(r_i)=\sqrt{U_{\rm exp}^2(r_i) + U_{\rm inv}^2(r_i)}, 
\end{equation}
where the first term is the experimental error coming from helioseismic frequencies, while the second and dominant term is the so-called "statistical" error in the inversion procedure estimated by \cite{scilla1997}. As in all previous works, we assume that there are no correlations between sound speed determinations derived at different radial points.
 
Both model and observational sound speed errors are presented in more detail in Sect.~\ref{soundspeedpro}.

\section{Method and statistical procedure}
\label{sec:methods}

The statistical approach is based upon constructing a $\chi^2$ function that can be used a figure-of-merit for the quality of different solar models in reproducing the observables. We build this function by considering 34 different observable quantities: the neutrino fluxes $\rm{\Phi(^8B)}$ and $\rm{\Phi(^7Be)}$; the convective envelope properties $\rm{Y_S}$ and $\rm{R_{CZ}}$ and the sound speed determinations ${\rm c}_i\equiv c(r_i)$ for 30 different value of $r/{\rm R_\odot}$ where $r/{\rm R_\odot} < 0.80$. We exclude the sound speed profile for $r/{\rm R_\odot} > 0.80$ because the transport is convective in that region, the temperature gradient is adiabatic and then, the sound speed profiles of the models are not affected by changes in the solar interior and are in agreement with helioseismic results. For values close to $r/{\rm R_\odot} \approx 1.0$, the predicted sound speed deviates from the observations due to the so-called surface or non-adiabatic effects that are not properly accounted for in the models. However, solar structure in this region is determined by the surface properties (mainly surface temperature and gravity) and, again, not influenced by the properties of exotic particles.

The fractional difference $\delta Q$ between theoretical predictions and observational data is defined as
\begin{equation}
\delta Q=1-\frac{Q_{\rm th}}{Q_{\rm obs}}
\end{equation}
where $Q_{\rm th}$ and $Q_{\rm obs}$  denote the predicted and observed values respectively for the quantity $Q$. The differencies $\delta Q$ are affected by uncorrelated experimental errors $U_{Q}$ and by correlated theoretical uncertainties $C_{Q, I}$. Following the formulation of \cite{fogli2002}, $\chi^2$ can be calculated as:
\begin{equation}
\chi^2=\min_{\{\xi_I\}}\left[ \sum_Q \left( \frac{\delta Q - \sum_I{\xi_I C_{Q,I}}}{U_Q}\right)^2 + \sum_I \xi_I^2 \right].
\label{eq:xi}
\end{equation}
where the shifts $-\xi_I C_{Q,I}$ describe the effects of correlated errors. They give the corrections of theoretical predictions $Q_{\rm th}$ when the input parameters $I$ are varied by the fractional amount $\xi_I \delta I$. To normalize the effect of these corrections, a penalty $\sum_I{\xi_I^2}$ is introduced in the $\chi^2$. The values $\tilde{\xi}_I$ that minimize the $\chi^2$ are referred to as pulls of correlated error sources.  The quadratic sum $\chi^2_{\rm syst} = \sum_I \tilde{\xi}^2_I$ give the systematic error contribution to the $\chi^2$ while the distribution of the pulls gives information about tensions between input parameters \cite{fogli2002} in the models. We remark that this method is completely equivalent to the use of the covariance matrix in order to compute the $\chi^2$ function. The main advantage of this approach is that it makes it possible to study the individual contributions to $\chi^2$. For more details about the statistical method applied to solar models we refer the reader to \cite{villante2014}.

If we keep the composition of the Sun fixed, there is only one free parameter in our calculations (either $g_{10}$ or the product $\chi m$) and the energy loss is proportional to its squared value. However, we want to prevent that incorrect assumptions about the surface composition of the Sun bias our final results. For this reason, we take a conservative approach for constraining properties of exotic particles and extend the idea presented in \cite{villante2014} where solar models with arbitrary composition are confronted with observational data.
Following \cite{villante2014}, we indicate with $\left\{ z_j\right\}$ the heavy element admixture, expressed in terms of the quantities $z_j \equiv {\rm Z}_{j,\rm S}/{\rm X_{S}}$ where ${Z}_{j,\rm S}$ is the surface abundance of the $j$-element, ${\rm X_{S}}$ is that of hydrogen, and the index $j$ runs over metals. We group the relevant elements for solar model calculations in {\em volatiles} (i.e. C, N, O and Ne) and {\em refractories} (i.e. Mg, Si, S and Fe) and we assume that the abundances $z_j$ within each group vary by the same multiplicative factors $(1+\delta z_{\rm vol})$ and $(1+\delta z_{\rm ref})$ with respect to the values prescribed by the adopted reference solar composition (either GS98 or AGSS09). We compute the effects of a composition change by linear expansion in $\delta z_{\rm vol}$ and $\delta z_{\rm ref}$  with the logarithmic derivatives given in \cite{villante2014} and we superimpose composition effects to those produced by axions or hidden photons energy losses.
This approach has been shown to reproduce well the structural (temperature, density) changes in solar interiors when the solar composition is varied \cite{villante2014}. Because axion and hidden photons energy losses do not depend on the detailed metal composition of the Sun and the allowed range of energy losses is quite limited, this procedure is also valid in solar models including axions or hidden photons.

In this approach, there are three free parameters in theoretical predictions: one ($g_{10}$ or $\chi m$) is related to axions or hidden photon properties and the other two $\left(\delta z_{\rm vol},\delta z_{\rm ref}\right)$ are used to rescale from the reference solar models. Because we are not interested in constraining composition, we minimize with respect to $\left(\delta z_{\rm vol},\delta z_{\rm ref}\right)$, i.e. for each value of $g_{10}$ or $\chi m$, we choose the solar surface composition that leads to the best agreement with observational data.  As a result, we obtain the function 
\begin{equation}
\tilde{\chi}^2\left(f\right)=\min_{\{\delta z_{\rm j}\}} \left[ \chi^2\left(f, \delta z_{\rm vol},\delta z_{\rm ref} \right) \right]
\label{eq:chi2}
\end{equation}
where $f = g_{10},\,\chi m$, that can be used to constrain axion and hidden photon properties.  The best fit values for  $g_{10}$ and $\chi m$ are found by minimizing  this function and the obtained value $\tilde{\chi}^2_{\rm min}$ provides information on the goodness of the fit.  The allowed regions for $g_{10}$ and $\chi m$ at 1, 2 and 3$\sigma$ confidence level (CL) are determined by cutting at the values of the variable $\Delta  \chi^2 \equiv \tilde{\chi}^2  -  \tilde{\chi}^2_{\rm min} =$1, 4, and 9 respectively, as it is prescribed for a chi-squared function with one degree of freedom. 

As we shall see, the redundancy of the different experimental information included in our global analysis allows us to disentangle the peculiar effects introduced by exotic particles from those produced by arbitrary composition changes. 
%and/or induced by variation of input parameters for SSMs within their range of uncertainties (see sect.~\ref{Uncert}).  
By using this extremely conservative approach, we obtain restrictive bounds which are clearly more robust than those obtained in analyses that consider one observable quantity at a time (e.g. the $^{8}{\rm B}$ neutrino flux or one point of the sound speed profile, see e.g.~\cite{part-schlattl99,part-gondolo09,maeda13}).

\section{Results}
\label{sec:results}

\begin{figure}[t]
	\centering 
	\includegraphics[width=0.6\textwidth]{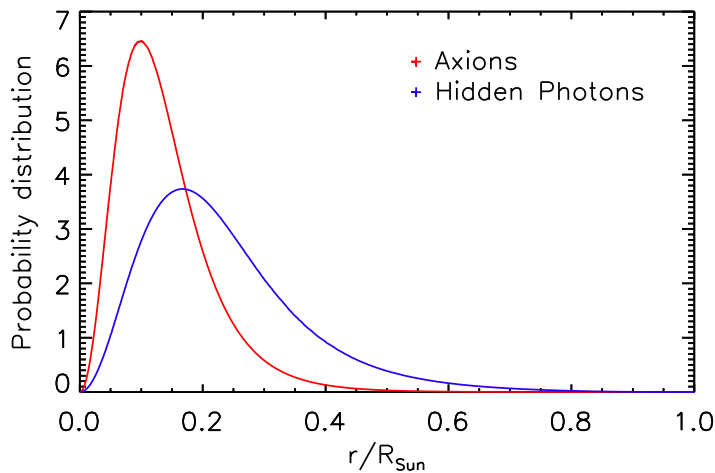}
        \caption{\small Normalized production distribution as a function of $r/R_\odot$. Red and blue lines correspond to axions hidden photons respectively.} \label{fig:emission}
\end{figure}

Equations~\ref{eq:primakoff}~and~\ref{eq:HP} give the dependence of the energy loss rates induced by axions and hidden photons on temperature and density. Different dependences translate into different changes in the structure of the Sun. To facilitate discussion and interpretation of results we show in Figure~\ref{fig:emission} the normalized production profile, i.e. the energy loss rate per unit mass multiplied by the mass contained in a spherical shell of radius $r$,  as a function of the solar radius for axions (red) and hidden photons (blue). The differences can be easily seen. The production of axions induces energy-losses in the inner region of the Sun ($r<0.4 {\rm R_\odot}$), whereas the energy-loss distribution is much broader for the case of hidden photons. The difference is mostly due to the temperature dependence of the energy-loss rates, see eqs. \ref{eq:primakoff} and \ref{eq:HP}. For axions, the rate $\epsilon_{a\gamma}$ is roughly proportional to $T^6$, while for hidden photons the dependence of the rate $\epsilon_{hp}$ on the temperature is approximately linear under solar conditions. As a rule of thumb, we expect that observables will be affected in a different way. For example, axions will have a major effect on the sound speed profile in the core regions while hidden photons can produce modifications of the sound speed profile in all regions. Notice, however, that although energy losses are localized in certain regions of the Sun, structural changes can be, to some degree, present all over the Sun. Indeed, local changes in certain quantities produce variations in the whole structure \cite{villante10}. 

Sections~\ref{sec:neutrconvec} and \ref{soundspeedpro} show the changes in the solar observables (Sect.~\ref{sec:observables}) induced by axions and hidden photons in models for which the solar composition is kept fixed at the reference values (GS98 or AGSS09). This makes it easy to compare the different response of solar models to different particles and to compare with previous work where the solar composition was not varied. However, as described before, the solar composition is free in the models from which we derive limits on particle properties. The best fit SSM obtained with this procedure is briefly presented in Sect.\,\ref{sec:bestssm}. Results of this exercise and the constraints we derive for both axions and hidden photons are described in \ref{sec:methods}.

\subsection{Solar neutrinos and convective envelope properties}
\label{sec:neutrconvec}

\begin{figure}
	\centering 
	\includegraphics[width=0.8\textwidth]{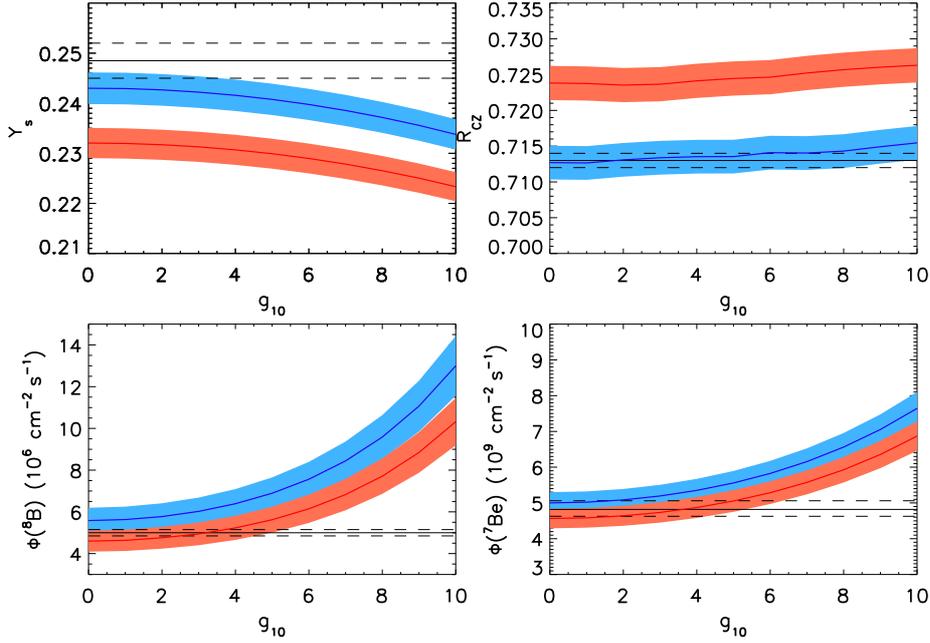}
        \caption{\small Evolution of the model parameters ($\rm{Ys}$, $\rm{R_{CZ}}$, $\rm{\Phi(^7Be)}$, $\rm{\Phi(^8B)}$) as a function of the axion-photon coupling constant. Red color correspond to AGSS09 abundances and the blue one to the GS98 abundances. Black lines represent the observational value with the errors and the shaded zones show model errors.}\label{fig:obsmodelax}
\end{figure}

\begin{figure}
	\centering 
	\includegraphics[width=0.8\textwidth]{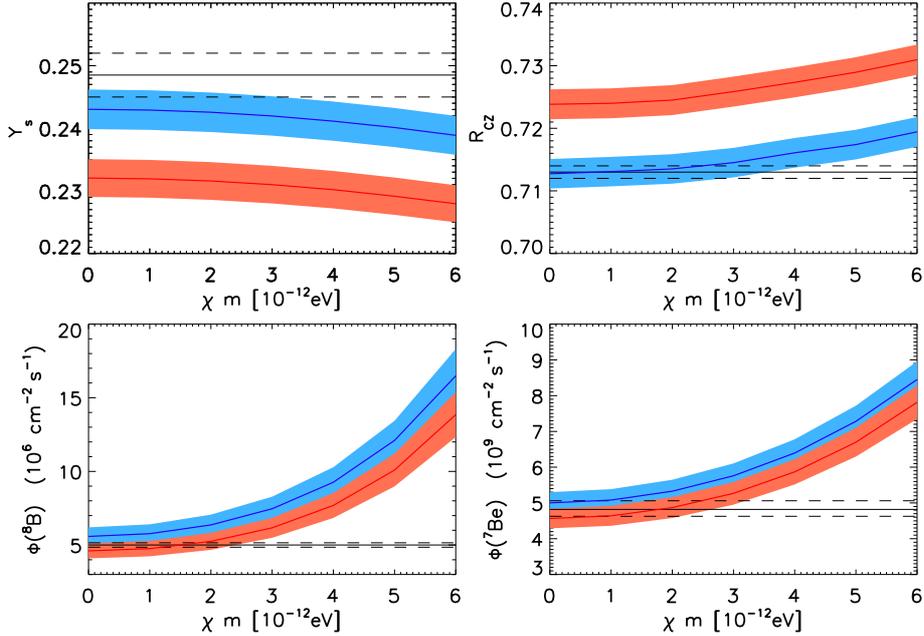}
        \caption{\small Same as Figure \ref{fig:obsmodelax} but for the hidden photons case. The variable on the x-axis corresponds to the product of the kinetic mixing and the hidden photon mass.} \label{fig:obsmodelhid}
\end{figure}

Figures \ref{fig:obsmodelax} and \ref{fig:obsmodelhid} show the dependence of the solar neutrino fluxes $\rm{\Phi(^8B)}$ and $\rm{\Phi(^7Be)}$ and of the convective envelope properties ${\rm Y_S}$ and ${\rm R_{CZ}}$ on $g_{10}$ and $\chi m$, respectively. Red lines correspond to solar models implementing AGSS09 composition and blue ones to GS98. The shaded zones depict the $1\sigma$ theoretical errors calculated as described in sec.~\ref{Uncert}. The black lines shows the experimental values with $1\sigma$ errors, as given in Table~\ref{tab:obs_value}.

%We note that the same quantitative effects are produced by axions and hidden photons on all these quantities (and also on the sound speed, as discussed in next section) irrespective of the assumed solar composition. This validates our approach in which we superimpose composition and exotic particle effects to calculate the properties of solar models that include axions or hidden photons and which have modified surface composition with respect to AGSS09 prescriptions.  

In the upper panels of Figure~\ref{fig:obsmodelax}, we see how the observables ${\rm Y_S}$ and ${\rm R_{CZ}}$ change in the presence of axions. The surface helium abundance ${\rm Y_S}$ decreases with increasing values of the axion-coupling constant because it is almost perfectly correlated to the initial helium abundance ${\rm Y_{ini}}$. This quantity decreases with $g_{10}$ because a higher initial amount of hydrogen is necessary to match the solar luminosity ${\rm L_\odot}$ in the presence of axion energy losses. The change in the convective radius ${\rm R_{CZ}}$ is quite small because energy losses are localized in the innermost regions of the Sun.

The lower panels present results for the neutrino fluxes. $\rm{\Phi(^8B)}$ and $\rm{\Phi(^7Be)}$  increase with increasing $g_{10}$ predominantly as a result of higher core temperatures and reach values well outside of the theoretical and experimental 1-$\sigma$ errors already at  relatively small values of  $g_{10}$. The $\rm{\Phi(^8B)}$ relative changes are larger than those of $\rm{\Phi(^7Be)}$ due to its stronger sensitivity to temperature. As a consequence, we expect that $^8{\rm B}$ neutrino measurements give stronger constraints on $g_{10}$. 
%We note that $\rm{\Phi(^8B)}$ alone has already been used to place constraints on $g_{a\gamma}$ \cite{part-gondolo09,maeda13}. 
It is important to mention that the present experimental determinations of  $\rm{\Phi(^8B)}$ and $\rm{\Phi(^7Be)}$ have very small uncertainties, 3\% and 4.5\% respectively. The constraining power of both fluxes, and particularly $\rm{\Phi(^8B)}$, is currently limited by uncertainties in solar models. 

Figure~\ref{fig:obsmodelax} also illustrates that the relative variations in the observables do not depend on the reference solar composition used, as expected because axion production does not depend directly on it. Differences between models with GS98 and AGSS09 compositions simply reflect the initial differences in the respective SSMs.

In the case of hidden photons, the changes of the solar neutrino fluxes and convective envelope properties have the same qualitative behavior, as it is seen in Figure~\ref{fig:obsmodelhid}. The decrease of ${\rm Y_S}$ and the increase of $\rm{\Phi(^7Be)}$ and $\rm{\Phi(^8B)}$ are again a consequence of the solar luminosity constraint, i.e. of the fact that increased initial hydrogen abundance and increased core temperature are necessary to compensate energy losses. 
The behaviour of ${\rm R_{CZ}}$ is instructive because it shows that different types of particles can induce different changes in the solar structure. Let us consider, e.g., models for axions and hidden photons for which the boron neutrino flux is comparable, say $\rm{\Phi(^8B)}\sim 15\cdot10^6\hbox{cm$^2$s$^{-1}$}$. In the case of axions, this corresponds to solar models with an increase of the convective radius equal to $\Delta {\rm R_{CZ}}\approx 0.003{\rm R_\odot}$. For hidden photons, the increase of  ${\rm R_{CZ}}$ is a factor of two larger, i.e. $\Delta {\rm R_{CZ}}\approx 0.006{\rm R_\odot}$. For a given change in the central conditions, as essentially determined by $\rm{\Phi(^8B)}$, hidden photons lead to larger changes than axions in the outer layers as a consequence of the milder temperature dependence of their emission rate. 

\begin{figure}
\centering
\includegraphics[width=0.6\textwidth]{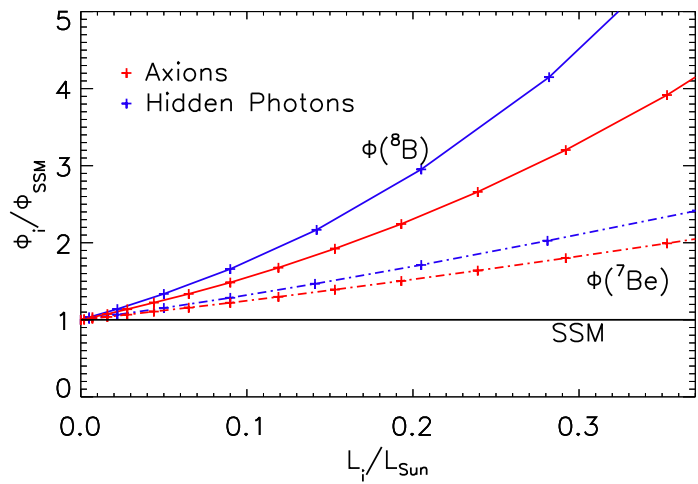}
\caption{\small Relative changes of the neutrino fluxes with respect to SSM prediction as function of the luminosity contribution of axions (red line) and hidden photons (blue line).}
\label{fig:discussion2}
\end{figure}

A direct comparison between results of Figures~\ref{fig:obsmodelax} and \ref{fig:obsmodelhid} should not be done because they are shown as functions of parameters not related to each other in a direct way. To compare the effects of both particles, Figure~\ref{fig:discussion2} illustrates the fractional change of the neutrino fluxes with respect to the SSMs for axions (red lines) and hidden photons (blue lines) as a function of the luminosity $L_i/{\rm L_\odot}$ carried away either by axions or hidden photons, i.e. $L_i=L_a$ or $L_{hp}$.

For the same $L_i/{\rm L_\odot}$ value, changes in the neutrino fluxes are larger for the hidden photons models. 
It was shown by \cite{part-gondolo09} that $L_a$ and $\rm{\Phi (^8B)}$ can be related by a simple analytic relation of the form
\begin{equation}
\frac{\rm{\Phi(^8B)}}{\rm{\Phi_{\rm SSM}(^8B)}} = \left(\frac{L_a + {\rm L_\odot}}{{\rm L_\odot}}\right)^\alpha, \label{eq:power}
\end{equation}
where $\alpha=4.6$, based on the older generation of SSMs computed by \cite{part-schlattl99}. Our calculations yield $\alpha= 4.4$ for axions, very close to the previous result. Interestingly, for hidden photons we find that the same functional form can be used but with a much steeper relation given by $\alpha= 5.7$.

These results reinforce the importance of performing self-consistent solar model calculations to account for the effect of exotic particles. Also, they show that assuming a universality in the constraints that are imposed, e.g. a given $L_i/{\rm L_\odot}$ value, can lead to biased bounds on the properties of particles.

\subsection{Solar sound speed profile}
\label{soundspeedpro}

\begin{figure}
\centering
\includegraphics[width=1.0\textwidth]{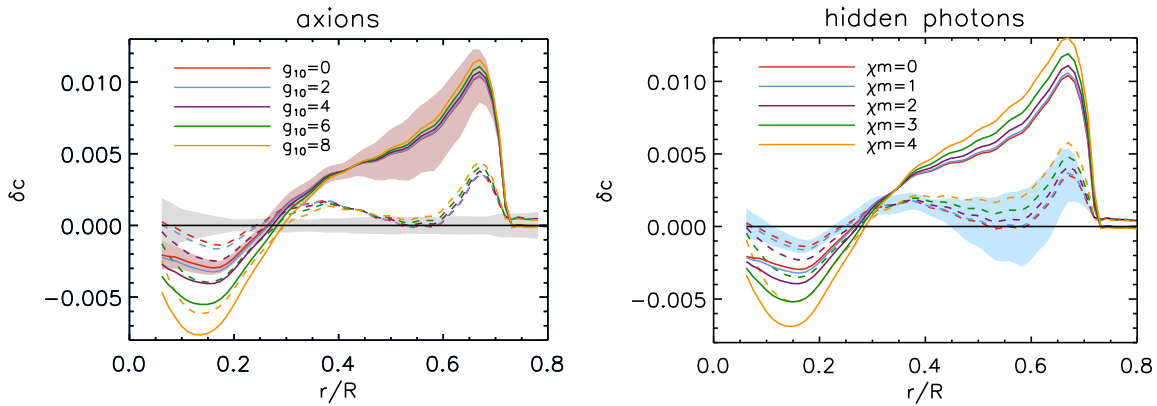}
\caption{\small \label{fig:cspro} 
Sound speed profile of models with axions (left panel) and hidden photons (right panels) for different values of the axion-photon coupling constant $g_{10}$ and of the product $\chi m$ (expressed in $10^{-12} \rm{eV}$) for hidden photons. Models are calibrated to the reference solar compositions GS98 (dashed lines) and AGSS09 (solid lines). Red and blue shaded zone corresponds to the model errors and the grey one to errors in the helioseismic inversion.}
\label{Fig:cs}
\end{figure}

In Figure~\ref{fig:cspro} we represent the sound speed profiles $\left(\delta {\rm c_s}= \frac{\rm c_{s, obs}-c_{s, th}}{\rm c_{s, obs}}\right)$ of models including axions (left panel) and hidden photons (right panel). For each case, five different values of $g_{10}$ and $\chi m$ are considered. SSMs correspond to the null value of each parameter. The solid lines are obtained for solar models implementing AGSS09 surface composition while the dotted lines correspond to GS98 surface admixture. The shaded zones show the theoretical uncertainties (blue and red colors; not including composition uncertainties) and the uncertainties coming from the inversion technique (grey color).

We see that axions produce effects on the sound speed in the inner region of the Sun.
This is in line with the results obtained for the convective radius ${\rm R_{CZ}}$, which showed little variation in the presence of axion energy losses. This implies that the constraints on $g_{10}$ would be obtained from the measurements of the sound speed at $r/R_\odot < 0.35$. On the other hand, hidden photons produce noticeable effect also in more external regions, meaning that the entire sound speed profile could potentially contribute to constrain the product $\chi m$. However, this is not so simple because in our statistical method composition is free to vary. Variations in the sound speed profile will be compensated, at least partially, by changes in the solar composition.

As it is seen from Figure~\ref{fig:cspro}, solar models implementing AGSS09 surface admixture provide a poor description of the sound speed profile inferred from helioseismic data. In particular, the sound speed prediction deviates at the bottom of the convective envelope by about $\sim1\%$ with the helioseismic values. In this region, the modifications introduced either by axion or hidden photon energy losses are generally small and cannot explain the observed discrepancy. On the contrary, for $r/{\rm R_\odot} < 0.35$, where axion or hidden photons effects are most relevant, the sound speed profiles of models with fixed reference surface composition deviate in the opposite direction as would be required to resolve the abundance problem. 

Although the composition is kept fixed in models shown in Figure~\ref{fig:cspro}, results suggest that the information encoded in the solar sound speed profile should be able to help constraining the axion and hidden photon properties.

\subsection{Best standard solar model}\label{sec:bestssm}

\begin{figure}
\centering
\includegraphics[width=0.6\textwidth]{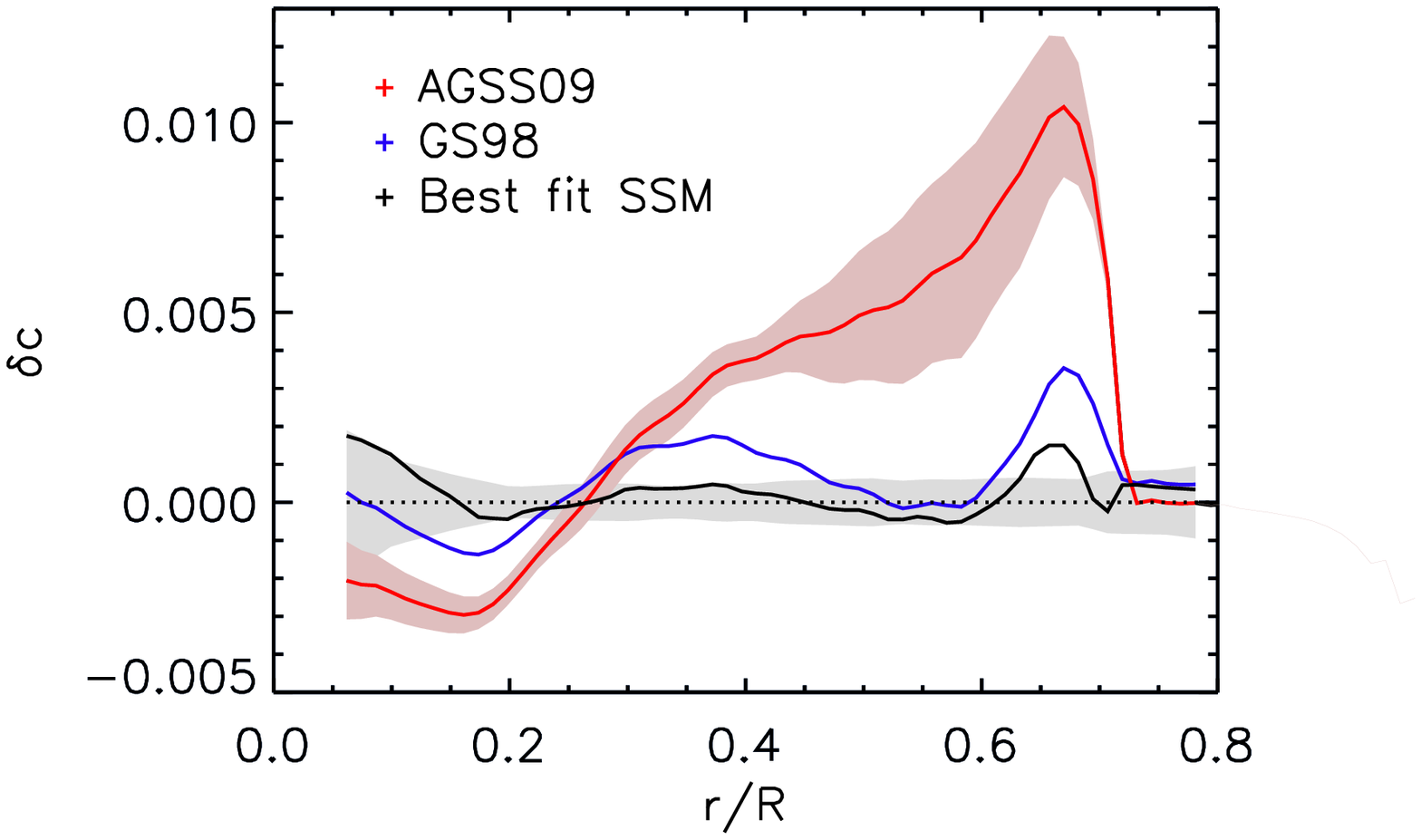}
\caption{\small \label{fig:csprossm} Comparison of sound speed profiles of SSMs. Red and blue lines are SSMs with AGSS09 and GS98 reference compositions and all input SSM parameters fixed to their central values. The black line shows results for the best SSM \cite{villante2014} resulting from finding the SSM with free composition that satisfies Eq.\,\ref{eq:chi2}. }
\end{figure}

Our statistical approach derives from that in \cite{villante2014}, where it was applied to SSMs. By letting the solar composition free and and by adjusting the input parameters in SSMs (nuclear cross sections, microsocopic diffusion rate, etc.) thorugh the pulls $\xi_I$ (that contributes to the total $\chi^2$ through the penalty factor $\sum_I \xi_I^2$), it is possible to find the SSM that best reproduces helioseismic and solar neutrino data. Details on the calculations and results are given in \cite{villante2014}. Here, we just present a quick overview. In Figure~\ref{fig:csprossm} we compare the sound speed profiles of the two SSMs with GS98 and AGSS09 solar compositions and the best SSM (best fit SSM) resulting from \cite{villante2014}.

This figure illustrates the improvements of the best fit SSM over both the AGSS09 and even the GS98 SSM that describes heliseismic properties quite well. In particular, the peak in the sound speed profile right below the convective zone at $r/{\rm R_\odot}=0.713$ is now smaller than 0.15\% and the behavior of the sound speed in intermediate regions between $r/{\rm R_\odot}=0.5$\,and\,0.2 is also greatly improved. The best fit SSM is achieved  with composition parameters $\delta z_{\rm vol} = 0.45$ and $\delta z_{\rm ref} = 0.19$ that correspond to the logarithmic astronomical abundances $\varepsilon_{\rm O} = 8.85$ and $\varepsilon_{\rm Fe} = 7.52$ for oxygen and iron respectively (the most relevant elements among volatiles and refractories, respectively), very close to GS98 values. The systematic pulls generally indicate variations at a  1-$\sigma$ level or less from the central values for the input parameters of solar models. The most relevant systematic pulls of the best fit SSM are: $\xi_{\rm opa}=1.07$, $\xi_{\rm diff}= -0.50$, $\xi_{\rm S34}= -1.2$, $\xi_{\rm S17}= -1.30$. For this model, $\chi^2_{\rm min} = 38.5$ for  34 observational constraints (see Sect.\,\ref{sec:observables}) indicating a good agreement between the model and the data. For comparison, the $\chi^2$ of  the SSM is  $179.6$ when the composition is kept fixed at the AGSS09 values and 62.4 when fixed to GS98.
%\footnote{In order to compare with $\chi^2_{\rm min}=38.5$, the quoted $\chi^2$ value is obtained not including the chemical composition contribution to the total error budget. For completeness, the $\chi^2$ value obtained by including the AGSS09 composition uncertainties is $72.5$.}.}

The best fit model is almost completely independent on the reference solar composition that is initially used. This is partly due to the fact that the coefficients $B_{Q,I}$ (eq.\,\ref{eq:bqi}) are model independent to a very good approximation. Also, because variations between AGSS09 and GS98 compositions are very similar for all elements grouped as volatiles ($\sim$\,0.14\,dex) and then for those grouped as refractories ($\sim$\,0.05\,dex). Therefore, the two-parameter analysis performed in \cite{villante2014}, and upon which we base the analysis in this paper, can be considered as a family of solar compositions that smoothly connects the two GS98 and AGSS09 SSMs. The reference solar composition, then, becomes irrelevant for our analysis because the problems associated to the solar abundance problem can be efficiently compensated by the varying the composition and the systematic pulls.

The best fit SSM represents our best effort to reproduce helioseismic and solar neutrino data. Solar models constructed to reproduce seismic data have been obtained previously by other authors. A couple of examples are the so-called solar seismic model \cite{turck2001} and the solar model by \cite{maeda13}. However, those models are not fully consistent with evolutionary history of the Sun because they are constructed as snapshots that reproduce present day structure. Our approach reproduces the present day structure consistently with evolutionary history as well.

As a final comment, it should be added that the description of the solar properties that the best fit SSM offers should be understood as a phenomenological description of the solar interior structure. Alternatively, one could let the opacity profile free and constrain the composition (e.g. to AGSS09 values) and recover the same level of agreement between solar data and models \cite{cd2009,villante10,villante2014}.

\vskip.5cm
\subsection{Global results}

\label{sec:global}

In this section we present the results of our global analysis. The bounds on axions and hidden photons are obtained by marginalizing with respect to surface composition, i.e. for each assumed value of for $g_{10}$ and $\chi m$ we rescale the surface abundances of volatile and refractory elements by the factors $\left(1+\delta z_{\rm vol}\right)$ and $\left(1+\delta z_{\rm met}\right)$ in order to achieve the best possible agreement with observational data. The null values of the chemical composition parameters $\delta z_{\rm vol}$ and $\delta z_{\rm met}$ correspond to the GS98 or AGSS09 photospheric composition (depending on the composition used to calculate the reference solar model). As expressed before, results have a very minimal dependence on the reference solar composition used. For simplicity, we show here the results obtained by using the AGSS09 as reference composition (i.e. as pivot point for expansion in $\delta z_{\rm ref}$ and $\delta z_{\rm vol})$. Identical results are obtained if GS98 composition is instead used. 

\begin{figure}[t]
\centering 
\includegraphics[width=0.9\textwidth]{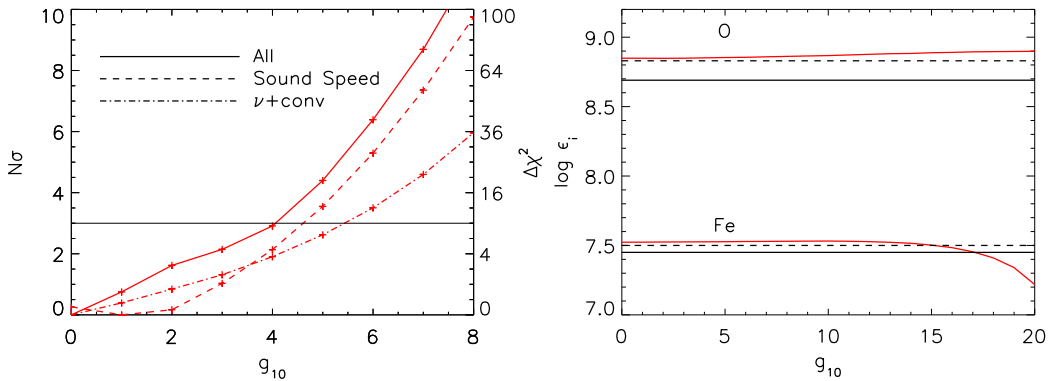}
\caption{\small \label{fig:chiallax} \textit{Left panel}: values of $N \sigma$ and $\Delta \chi^2$ for models with axions. Solid line: using all observables $\rm{\Phi(^7Be)}$, $\rm{\Phi(^8B)}$, $\rm{Y_s}$, $\rm{R_{CZ}}$ and 30 points of the sound speed profile. Dashed line: using the sound speed. Dotted-dashed line: using the neutrinos and convective envelope properties. \textit{Right panel}: best fit composition as a function of \gd\  presented as the logarithmic astronomical abundances (red solid lines). Black lines represent the GS98 (dashed) and AGSS09 (solid) values for $\epsilon _O$ and $\epsilon _{Fe}$.}
\end{figure}

\emph{Axions -}  There is no model including axions ($g_{10} \neq 0$) that improves the overall fit to the data with respect to the our best fit SSM (Sect.\,\ref{sec:bestssm}).
The variation of $\chi^2$ and its equivalent $N\sigma=\sqrt{\Delta \chi^2}$ as a function of $g_{10}$ are shown with solid line in left panel of Fig.\,\ref{fig:chiallax}. The right panel shows the values of the logarithmic abundances $\varepsilon_{\rm O}$ and $\varepsilon_{\rm Fe}$ that provides the best fit to the data as a function of ${g_{10}}$. These quantities are almost independent on ${g_{10}}$ indicating that there are no degeneracies between composition and axion effects. Moreover, the best fit composition is very close to GS98 values. This should not be intended as a proof that the AGSS09 admixture is wrong but more as an evidence that the thermal stratification of SSM implementing GS98 composition is a good approximation of that of the real Sun.  

By setting a limit at $\Delta \chi^2=9$, we derive the upper bound $g_{10} < 4.1$  at a 3-$\sigma$ CL. This is almost a factor of 2 lower than previous solar limits (Sect.\,\ref{sec:discussion}).

%PULLS
One of the advantages on using equation \ref{eq:xi} to calculate $\chi^2$ is the possibility to understand which are the contributions to the total value of $\chi^2$. As it has been discussed in \cite{fogli2002} and \cite{villante2014}, the method  allows us to separate $\chi^2_{\rm{obs}}$ and $\chi^2_{\rm{sys}}$. If we compare those values for the SSM and for the model with ${g_{10}}=4$, we find that for the SSM $\chi^2_{\rm{obs}}=33.6$ and $\chi^2_{\rm{sys}}=4.9$ while for the model with ${g_{10}}=4$ we obtain $\chi^2_{\rm{obs}}=35.3$ and $\chi^2_{\rm{sys}}=11.7$. The values for $\chi^2_{\rm{obs}}$ are similar for both cases. This is the result of letting the composition to adjust to provide a good fit the solar data. Also, the changes that axions induce in solar models (see Sects.\,\ref{sec:neutrconvec}\,and\,\ref{soundspeedpro}) are partially compensated by the systematic pulls. But changes to the input parameters  come at the expense of increasing $\chi^2_{\rm{sys}}$, that thus provides the dominant contribution to $\Delta \chi^2$. The dominant pulls that increase the value of the $\chi^2$ with increasing $g_{10}$ are the solar luminosity (decrease), $\rm{S_{11}}$ (increase) and, to a lesser extent, $\rm{S_{17}}$ (decrease). The first two are mostly related to changes in the sound speed in the solar core. A lower solar luminosity and a larger ${\rm S_{11}}$ both contribute to decrease the theoretical sound speed in the solar core \cite{schlattl99b}, compensating the effect of axions (Fig.\,\ref{Fig:cs}). Changes in ${\rm S_{17}}$ occur in order to limit the increase in $\Phi({\rm ^8B})$ shown in Fig.\,\ref{fig:obsmodelax}.

It is also instructive to discuss how much each piece of experimental information contributes to the bound on $g_{10}$. To this aim, we show in the left panel of Fig.\,\ref{fig:chiallax} with dotted lines the $\Delta \chi^2$ functions obtained by considering separately the sound speed profile on one hand, and the neutrino fluxes 
$\rm{\Phi(^8B)}$ and $\rm{\Phi(^7Be)}$ and convective envelope properties $\rm{Y_S}$ and $\rm{R_{CZ}}$ on the other. We refer to this last  dataset with the label \texttt{$\nu$+conv} in the following. The most restrictive limit comes from the sound speed profile that, alone, sets the limit $g_{10}<4.6$ at $3\sigma$ CL. We note, however, that also the \texttt{$\nu$+conv}  dataset provides a restrictive bound $g_{10}<5.5$ at $3\sigma$ CL. The latter value is more restrictive than found \cite{part-gondolo09} using only the $\rm{\Phi(^8B)}$ even if composition is free in our fit. This result that may seem surprising can be explained from the results presented in \cite{villante2014}. It was indeed shown in that paper that the two observable quantities $\rm{Y_S}$ and $\rm{R_{CZ}}$ permit to determine the surface composition of the Sun (in the two parameter analysis in terms of $\delta z_{\rm vol}$ and $\delta z_{\rm met}$) with good accuracy. The experimental information on the neutrino fluxes, which are strongly dependent on $g_{10}$, can be then effectively translated into a bound on axion energy losses.

\begin{figure}[h!]
\centering 
\includegraphics[width=0.6\textwidth]{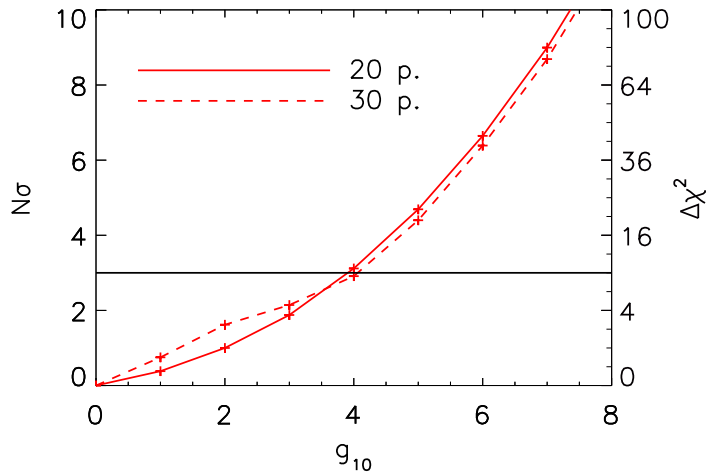}
\caption{\small \label{fig:20p} The values of $N \sigma$ and $\Delta \chi^2$ for models with axions using $\rm{\Phi(^7Be)}$, $\rm{\Phi(^8B)}$, ${\rm Y_s}$, $\rm{R_{CZ}}$ and 20 points of the sound speed profile as input observables.}
\end{figure}

Finally, since the sound speed profile provide the most restrictive constraint, we check that the bump of ${\rm c_s(r)}$ observed in the tachocline, i.e. the region just below the convective boundary, does not affect the final results in a critical way. This bump is present even in models with optimized composition (see e.g. the sound speed profile of the best fit SSM in Fig.~\ref{fig:csprossm}) and is due to the inadequacies of SSMs in modeling dynamic effects in that region that affect the composition gradient and the transition between adiabatic and radiative gradient. (e.g. \cite{gough96,basu97,bump2011}). We repeated the global analysis excluding the sound speed data profile in the region $r/R_\odot > 0.6$, thus reducing the sound speed data set to 20 data points. The exclusion of  the sound speed determinations in the external radiative region, as it expected, improves the quality of the fit being $\chi^2_{\rm min} =12.4$ for  24 observational constraints. However, since the limits are derived from the $\Delta \chi^2$ distribution, the bound for $g_{10}$ does not significantly change. In fact, by excluding this regions, as can be seen in Figure~\ref{fig:20p}, $g_{10}<3.9$ at $3\sigma$ CL which is an even more stringent limit than the case where the full sound speed profile (in combination with the other experimental informations) is used. The fact that axions mainly affect the sound speed in the inner region of the Sun (see Figure~\ref{fig:cspro}) explains why the results depend only weakly on the inclusion of the bump. And, using the full sound speed the bound of $g_{10}$ is more conservative.

\begin{figure}[h]
\begin{center}
\includegraphics[width=0.6\textwidth]{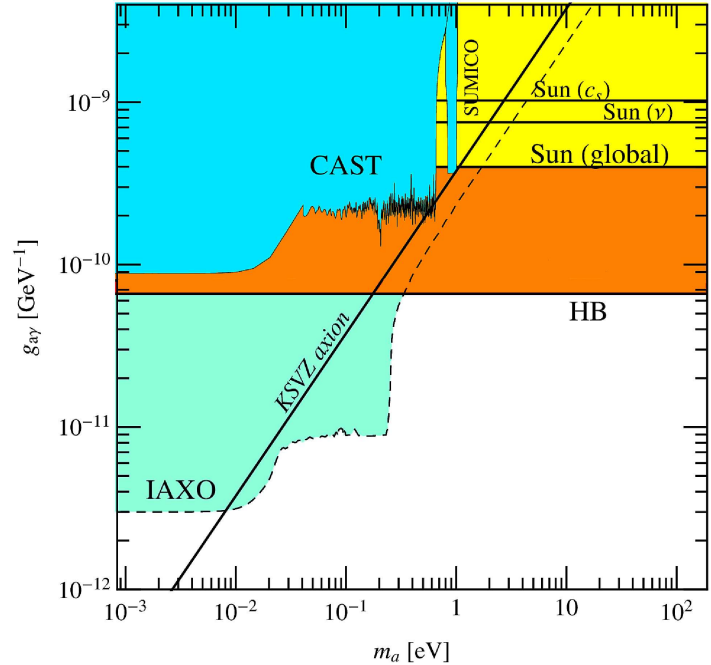}
\caption{\small Constraints on axion-like particles with a two-photon coupling as a function of the mass. 
The hadronic axion would be a point in the KSVZ line. The global fit presented in this paper provides a constraint (Sun (global)) that improves over the solar constraint obtained only with the solar neutrino argument (Sun $\nu$) and previous seismology constraints (Sun $c_s$) by a significant factor. 
Still, more stringent bounds come from the lifetime of Horizontal Branch stars in globular clusters. For masses below $\sim$eV the axion heliosocopes CAST and SUMICO provide very competitive limits and the future International Axion Observatory (IAXO) has the potential of improving over them all. }
\label{gagplot}
\end{center}
\end{figure}

Considering results discussed in this section, our recommended upper limit is $g_{a\gamma}\,<\,4.1\,\cdot\,10^{-10}\,\rm{GeV^{-1}}$ at 3-$\sigma$ CL for the axion-photon coupling constant. In Fig.\,\ref{gagplot} we summarize the relevant astrophysical constraints for hadronic KSVZ axions and axion-like particles, including our newly derived limit, together with the prospects of the future IAXO.

\begin{figure}[h!]
\centering
\includegraphics[width=1.\textwidth]{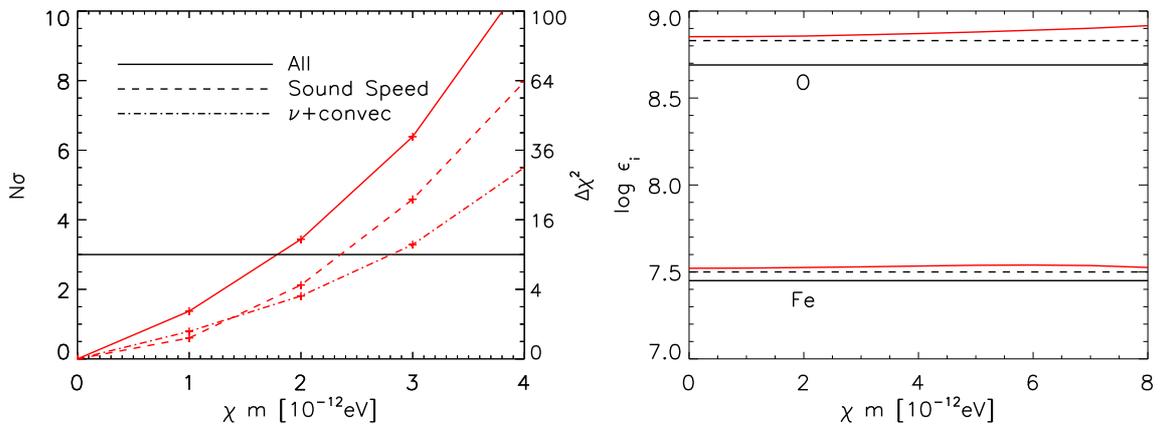}
\caption{\small Same than figure \label{fig:chiall} but for models with hidden photons.}
\label{fig:chiallhd}
\end{figure}

\vskip .5cm

\emph{Hidden photons -} In Figure~\ref{fig:chiallhd}, we show the results of our analysis for models including hidden photons. The left panel shows, the $\Delta \chi^2$ distribution and the corresponding $N \sigma$ values as a function of the product $\chi m$, as obtained by using all observational constraints. The right panel shows the values of $\varepsilon_{\rm O}$ and $\varepsilon_{\rm Fe}$ that minimize the $\chi^2$ for each assumed values of $\chi m$. In this case, as for the axion models, the best fit corresponds to the SSM with $\chi m=0$ and the volatile and refractory abundances increased by $\delta z_{\rm vol} = 0.45$ and $\delta z_{\rm ref} = 0.19$ with respect to AGSS09 composition that translate to  $\varepsilon_{\rm O}=8.85$ and $\varepsilon_{\rm Fe}=7.52$. 
The bound at the $3\sigma$ CL is given by $\chi m< 1.8 \cdot 10^{-12} \rm{eV}$ when the complete sound speed profile is used. If the region $r > 0.6\, {\rm R_\odot}$ is excluded, then the limit is only marginally different, $\chi m<1.7 \cdot 10^{-12} \rm{eV}$. Here again, using the full sound speed profile gives a more conservative limit than excluding the sound speed bump at the base of the convective envelope.

tThe different observational data contributes to the final result as in the axion case. The sound speed is the most restrictive observable, giving the constraint $\chi m < 2.4 \cdot 10^{-12} \rm{eV}$ at 3$\sigma$ CL. The neutrinos fluxes and convective envelope properties also give the relevant constraint $\chi m < 2.8 \cdot 10^{-12} {\rm eV} $ at $3\sigma$ CL.

\begin{figure}[h]
\begin{center}
\includegraphics[width=10cm]{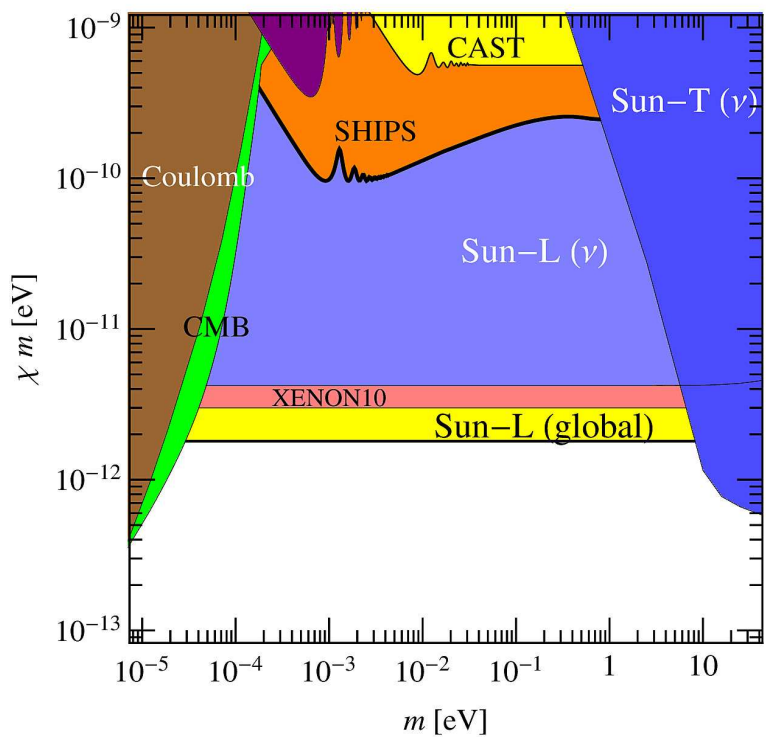}
\caption{\small Constraints on hidden photons with mass $m$ and kinetic mixing with photons $\chi$. 
The global fit presented in this paper, Sun-L (global),  provides a constraint  that improves over the solar constraint obtained only with the solar neutrino argument, Sun-L ($\nu$), and the direct detection constraint from XENON10 data~\cite{An:2013yua}.}
\label{HPplot}
\end{center}
\end{figure}

As a final result of our analysis, we quote $\chi m \le 1.8 \cdot 10^{-12} \rm{eV}$ at 3-$\sigma$ CL as an upper bound to product of the kinetic mixing parameter and mass of hidden photons. This limit does not depend on the assumed solar surface composition. In Fig.~\ref{HPplot} we discuss our new constraint in the mass-mixing plane together with other concurrent limits. Notably, the constraint is the most stringent in the range from $3\cdot 10^{-5}$ eV to 8 eV.

\section{Discussion}
\label{sec:discussion}

Here we have revised the solar limits on the axion-photon coupling constant and the product of the kinetic mixing and mass of the hidden photons by combining different solar constraints. We have used a statistical approach that accounts for both experimental and theoretical errors and lets the composition free in order that the solar abundance problem do not bias our final result. 

Previous works on the subject have often relied on setting an upper limit to the total energy loss carried away from the Sun expressed as a fraction of the solar photon luminosity ${\rm L_\odot}$. It is then useful to present our results as a function of $L_i/{\rm L_\odot}$. Also, $L_i/{\rm L_\odot}$ sets a natural scale for comparing the effects of axion and hidden photon losses to each other.

\begin{figure}[h!]
\centering
\includegraphics[width=0.6\textwidth]{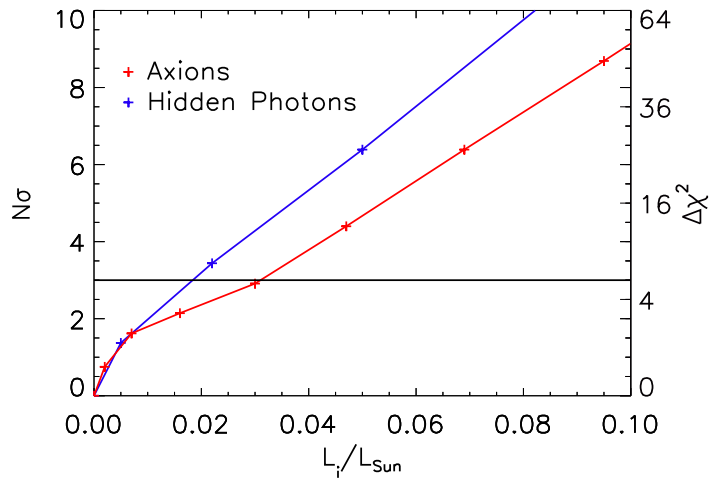}
\caption{\small $\Delta \chi ^2$ and $N\sigma$ as function of the luminosity contribution of axions (red line) and hidden photons (blue line).}
\label{fig:discussion1}
\end{figure}

In Figure~\ref{fig:discussion1}, global results from the previous section are presented again (only for the AGSS09 sets of models) but now as a function of $L_i/{L_\odot}$. We observe that the 3-$\sigma$ constraint for hidden photons corresponds to a contribution on the solar luminosity of only about 2\% while for the axions this value is around 3\%. For comparison, in \cite{part-gondolo09} the upper limit to axions is set by demanding $L_a < 0.1 {\rm L_\odot}$. We find that a much more restrictive limit can actually be imposed. 

It is interesting to note that a smaller fraction of ${\rm L_\odot}$ is required for the hidden photons than for axions to set a given confidence level. We can explain this by considering how the observables change in both cases. In Fig.~\ref{fig:cspro} we have seen that the changes on the solar sound speed profile are localized towards the center for the axion case while for the hidden photons the whole profile is affected.  Also, we find a steeper relation between the energy lost through hidden photons and the change in ${\rm \Phi(^8B)}$ than in the case of axions. All this contributes to placing somewhat stronger constraints for hidden photons than for axions in terms of the fraction of energy that is lost through these channels.

Solar limits on the coupling of hadronic axions to photons are not the most restrictive ones. However, this is a well-studied subject and it allows us to gauge the performance of our global statistical approach by comparing our findings to results from previous investigations. 

The first work that placed a limit on the axion-photon coupling constant using helioseismology is \cite{part-schlattl99}, where a set of solar models was calculated including effects of axions self-consistently. The bound $g_{10}<10$ was found by restricting the axion luminosity at $L_a < 0.2L_\odot$. For this limit, the deviations the sound speed profile are $0.8\%$ at $R=0.1R_\odot$. 

More recently, \cite{part-gondolo09} found $g_{10}<7$ by using the experimental $\rm{\Phi(^8B)}$ determination and, based upon its agreement with SSMs values, assuming that the actual solar flux cannot exceed the SSM prediction by more than 50\% (roughly equivalent to a limit of $3\sigma$ model uncertainty). As mentioned before, this is comparable to setting the limit $L_a < 0.1 {\rm L_\odot}$. 

A different approach has been followed in \cite{maeda13}, where a seismic model is used to determine the effect of the axions. 
Seismic models are static models of the Sun which, by construction, reproduce the sound speed profile.  They do not account for the evolutionary history which determines the chemical composition profile of the present Sun. Axions were included in the seismic models by adding the energy loss in the energy equation and modifying the central temperature and the abundances with respect the standard case so that the sound speed and solar luminosity are recovered\footnote{Note that, in this approach, it is not discussed whether the solar seismic model of the Sun with axions energy losses can be realized as the result of solar evolution.}. The constraint on $g_{10}$ is derived from the experimental value of $\Phi(^8{\rm B})$ and it is set to $g_{10} < 2.5$ at a $1\sigma$ CL. To compare, from Figure \ref{fig:chiall} we find a limit at a $1\sigma$ CL around $g_{10} < 1.5$. 

In comparison to these works, our result $g_{10}\,<\,4.1$ to a 3-$\sigma$ CL is much more restricting. This is a direct consequence of our global approach of combining consistently helioseismic and solar neutrino constraints. As discussed in the previous section, the sound speed profile is the most restrictive observational constraint. If we restrict our analysis to using \texttt{$\nu$+conv} , we obtain $g_{10}<5.5$ and a corresponding  $L_a < 0.06\,{\rm L_\odot}$ limit on the axion luminosity.

Using the limits on the parameters we have found for axions and hidden photons, we present upper limits for the respective fluxes on Earth expected in direct detection experiments such as CAST or IAXO. Using Eq.~15 from \cite{andriamonje2007} and our limit $g_{10}=4.1$ we obtain $\Phi_a \sim 6.0 \cdot 10^{12} \rm{cm^{-1} \cdot s^{-1}}$.  During the CAST data taking, the limiting flux has been found to be more restrictive for a wide range of axion masses. \cite{andriamonje2007} find a limit for the axion-photon coupling constant of $g_{10}=0.88$ for $m_a \lesssim 0.2 \rm{eV}$ corresponding to 
$\Phi_a \sim 2.9 \cdot 10^{11} \rm{cm^{-2} \cdot s^{-1}}$ and in \cite{arik2009} they find an upper limit of $g_{10}= 2.17$ for the mass range of $0.02<m_a<0.39 \rm{eV}$ corresponding to a solar flux of $\Phi_a \sim 1.8 \cdot 10^{12} \rm{cm^{-2} \cdot s^{-1}}$. As already stated in the introduction, our limit does not improve the best limits on $g_{a\gamma}$ but we can confirm that, at the CAST limits, it is not expected that axions would have a measurable effect with helioseismology and solar neutrinos.

Hidden photons have a younger history than axions in the literature, and previous bounds on the kinetic mixing parameter based on solar models are limited to \cite{an2013,solaredondo2013}. In the first case, a very conservative limit was derived by assuming that $L_{hp} < {\rm L_\odot}$, leading to $\chi m < 1.4 \cdot 10^{-11}{\rm eV}$. In the second, the more restrictive upper bound $\chi m < 4 \cdot 10^{-12}{\rm eV}$ for masses smaller than $m < 0.3 \rm{keV}$ was derived from the condition $L_{hp} < 0.1\,{\rm L_\odot}$ and Equation~\ref{eq:power}. We have improved this limit by including the sound speed profile in the analysis and a more consistent treatment of uncertainties in our global approach. As discussed above, the limit $\chi m < 1.8 \cdot 10^{-12}{\rm eV}$ is obtained from models for which $L_{hp}\,<\,0.02\,{\rm L_\odot}$, a much smaller fraction than employed in previous works.
Using Eqs~4.10~and~4.11 from~\cite{solaredondo2013}, the upper limit for the flux on the Earth we derive is $\Phi_{\rm{HP}} \sim 3.2 \cdot 10^{14} \rm{cm^{-2} \cdot s^{-1}}$ corresponding to $\chi m < 1.8 \cdot 10^{-12} \rm{eV}$.

We remark that our bounds on axion-photon coupling and hidden photons kinetic mixing from are not affected by the ongoing solar abundance problem, a crisis brought about by the latest generation of spectroscopic determinations of the solar photosphere composition. Indeed, we systematically studied the role of the composition used in solar models for the constraints on $g_{a\gamma}$ and $\chi m$ by letting the composition free in the study and marginalizing it for each value of \gd\  and $\chi m$. The results showed that the values for the composition are more or less constant with increasing \gd\  and $\chi m$ and that the values are close to the GS98 and the best fit SSM, implying that models with this composition are a good representation of the actual thermal stratification of the Sun. This is reflected in the values of $\chi_{\rm obs}$, that are almost invariant at the \gd\ or $\chi m$ 3-$\sigma$ limits compared to the best fit SSM. 

As shown in the literature \cite{cd2009,villante10,villante2014}, an equally good best fit SSM could be achieved by letting the radiative opacity vary instead of the solar composition. Keeping this in mind one concludes then that our limits are, to a good approximation, independent of the composition. This also indicates that exotic energy losses, even assuming the broad radial distribution produced by hidden photons, cannot be advocated as a possible explanation of the disagreement between solar models with AGSS09 composition and helioseismic data. 

There are different ways in which to move forward. Our global approach represents a qualitative step forward in combining different sources of solar data for using the Sun as a laboratory for particle physics. But further data is available which we have not considered. For example, frequencies of low-degree oscillations can be combined in an advantageous manner to gain further insight on the solar core (e.g. \cite{roxburgh2003,bison07}). Using these data can enhance the constraining power of helioseismology. The caveat is that correlations among different helioseismic observables have to be taken into account. We have already acknowledged that even the radial profile of the sound speed presents correlations that have not yet been quantified in the literature. This is the next step we will pursue for improving our statistical approach.

It is important to notice that current uncertainties are dominated by model uncertainties. Moreover, there is not a unique dominant error source. For example, the 11\% error in the theoretical $\Phi(^8{\rm B})$ results from 8\% error in $S_{17}$ and 7\% from radiative opacities. If one includes solar composition as a source of uncertainty, this contributes a further 9\%. Thus, reducing modeling errors is a difficult task. A similar situation happens for the sound speed profile, where the solar composition and the radiative opacities play comparable roles. 

On the other hand, the $\Phi(pp)$ and $\Phi(pep)$ fluxes have the interesting property that their theoretical errors are small, 0.6\% and 1.2\% respectively and therefore experimental determinations would place constraints on non-standard energy losses almost completely independent on solar modeling uncertainties. Recently \cite{borexino2014}, Borexino has directly measured the solar $\Phi(pp)$ for the first time, with an observational error of $\sim 10 \%$ and the solar $\Phi(pep)$ \cite{bellini2012} with an observational error of $\sim 20 \%$. Due to these large experimental errors, we have not included these fluxes as observational inputs because they would not contribute to the derived limits. In a future generation of experiments, reaching a 1\% error in such measurements could provide a cross comparison between the solar photon and neutrino luminosities. Considering the case of hidden photons, for which the solar constraint is the most restrictive one over a wide range of photon masses, such a measurement would translate into a hard limit $\chi m < 3 
\cdot 10^{-12}{\rm eV}$ at 3-$\sigma$ CL, provided the measured value agrees with SSM results, and smaller if it does not. This limit, albeit larger than our estimate, has the attractive advantage that is practically independent of solar modeling.

\section{Summary}
\label{sec:summary}

We have extended the statistical approach presented \cite{villante2014} that combines in a consistent manner helioseismic and solar neutrino data for using the Sun for particle physics studies. In order to avoid that the {\em solar abundance problem} could bias the final result, we considered the surface abundances of the Sun as free parameters in our analysis.

As a test case we have applied the method to the well studied case of hadronic axions and showed that previous solar bounds on the axion-photon coupling can be improved with our methodology. We have derived a strong upper limit of $g_{a\gamma}=4.1 \cdot 10^{-10} \rm{GeV}$ at a $3\sigma$ CL, almost a factor of two better than previous results. As a further application, we have considered energy losses by hidden photons. This case is particularly interesting because the Sun offers the most restrictive limits over a wide range of photon masses. We have derived a new upper limit for the product of the  kinetic mixing parameter and mass $\chi m = 1.8 \cdot 10^{-12 } \rm{eV}$ for a hidden photon mass in the range $m \lesssim 0.3$ keV, more than a factor of 2 improvement over previous results \cite{solaredondo2013} and better than the direct detection constraint of XENON10~\cite{An:2013yua}.

By comparing in detail results from axions and hidden photons, we also conclude that relations between \emph{dark} luminosities and observables such as $\rm{\Phi(^8B)}$, e.g. Equation~\ref{eq:power}, depend on the type of particle under consideration and should be employed consistently to avoid biasing results. Finally, we have checked that including hidden photons in solar models with AGSS09 composition degrades the agreement between models and helioseismology, and thus does not help in mitigating the \emph{solar abundance problem}. 

The approach is of course general and will be extended to account for correlations among observables, a fact hitherto neglected in the literature, that will allow us to include additional helioseismic constraints.

\acknowledgments{We acknowledge funding support from ESP2013-41268-R (MINECO), 2014SGR-1458 (Generalitat de Catalunya) and the MICINN grant AYA2011- 24704. J. R. acknowledges support from the spanish nations through the RyC fellowship 2012-10597. We also want to acknowledge Michele Maltoni for discussion and clarification on the determination of solar neutrino fluxes from experimental data. }

\bibliography{hiddenaxion3,stellarWISP}

\end{document}